\documentclass[aip,graphicx]{revtex4-1} 

\usepackage{graphicx}
\usepackage{dcolumn}
\usepackage{bm}

\usepackage{amssymb}
\usepackage{amsmath,ulem}
\usepackage{amssymb}
\usepackage{color, colortbl,soul,xcolor}
\usepackage{fullpage}
\usepackage{enumerate}
\usepackage{setspace}

\newcommand{\beq}{\begin{equation}}
\newcommand{\eeq}{\end{equation}}
\newcommand{\barr}{\begin{array}}
\newcommand{\earr}{\end{array}}

\newcommand\bda{\boldsymbol{\alpha}} 
\newcommand\bdb{\boldsymbol{\beta}} 
\newcommand\bdxi{\boldsymbol{\xi}}

\newcommand{\beqs}{\begin{equation*}}
\newcommand{\eeqs}{\end{equation*}}
\newcommand{\beas}{\begin{eqnarray*}}
\newcommand{\eeas}{\end{eqnarray*}}
\newcommand{\bea}{\begin{eqnarray}}
\newcommand{\eea}{\end{eqnarray}}

\usepackage{ulem}

\begin{document}


\title{Analysis of micro-fluidic tweezers in the Stokes regime}



\author{Longhua Zhao}
 \email[]{longhua.zhao@case.edu}
 \affiliation{Department of Mathematics, Applied Mathematics and Statistics, Case Western Reserve University, Cleveland, Ohio 44106, USA}

\author{Li Zhang}
\email[]{lizhang@mae.cuhk.hk}
\affiliation{ Department of Mechanical and Automation Engineering
The Chinese University of Hong Kong,
Shatin NT, Hong Kong SAR, China}

\author{Yang Ding}
\email[]{dingyang@csrc.ac.cn}
\thanks{corresponding author}
\affiliation{Division of Mechanics, Beijing Computational Science Research Center, Beijing 100193, China}


\date{\today}
\setstretch{2} 

\begin{abstract}
Nanowire fluidic tweezers have been developed to gently and accurately capture, manipulate and deliver micro objects. The mechanism behind the capture and release process has not yet been well explained. Utilizing the method of regularized Stokeslet, we study a cylindrical nanowire tumbling and interacting with spherical particles in the Stokes regime. The capture phenomenon observed in experiments is reproduced and illustrated with the trajectories of micro-spheres and fluid tracers. The flow structure and the region of capture are quantitatively examined and compared for different sizes of particles,  various tumbling rates, and dimensions of the tweezers. We find that pure kinematic effects can explain the mechanism of capture and transport of particles. We further reveal the relation between the capture region and stagnation points in the displacement field  , i.e., the displacement for tracer particles in the moving frame within one rotation of the wire.
\end{abstract}


\maketitle 

\section{Introduction}
\label{sect:intro}

Due to  advances in miniature technologies and  demands from industrial and biomedical applications, microfluidic devices, such as valves, pumps, and  mixers  have been developed and studied recently.  More information can be found from the review paper by Stone \& Kim \cite{stone2001}  and the references therein.
 Tweezers, which capture, transport and deliver particles in the microfluids, have drawn attention. A wide range of methods have been developed to achieve the tweezers' trapping and transportation functions, utilizing forces based on optical \cite{Liu2016}, electrical \cite{Cohen2006Suppressing}, magnetic \cite{Lee2004Manipulation}, and acoustic \cite{li2013chip} potential fields, or simply fluidic forces \cite{Petit2012,Zhou2017Dumbbell,Karimi2013}. Due to the non-contact and gentle features, the fluidic tweezers are particularly promising for biomedical applications. To better design and utilize fluidic tweezers, it is essential to understand the interaction between a particle and the flow structure, which is a fundamental fluid mechanics problem \cite{hjelmfelt1966motion,kaftori1995particle}.

One of the fluid structures used to interact and to manipulate particles is the microfluidic vortex. By generating a vortex on a micro-scale, researchers have invented micro-fluidic devices for centrifuging \cite{mach2011automated}, mixing \cite{lee2010rapid},  particle sorting \cite{wang2013vortex}, and separating particles from fluids \cite{Haller2015Microfluidic}.  Using corners in channels is one way to generate vortices.  For example, Tanyeri \& Schroeder \cite{tanyeri2013manipulation} built a single particle manipulation and confinement device based on perpendicularly crossing channels. Such devices often utilize the inertial effect of the particle to deviate its trajectory from the fluid. Another way to generate vortices is to rotate an object controlled by an external magnetic field or ultrasound. These methods have been shown to be useful for propulsion and manipulation of particles  \cite{balk2014kilohertz,Petit2012}. 
Qualitative force analysis has been done for  rotational tweezers by Ye {\it et al.} \cite{ye2012micro,ye2014dynamic}. However, the mechanism for capture has not been fully explained. 
Although inertial effect and non-Newtonian effects have been considered as the mechanisms for trapping in the previous studies, and such effects could be responsible for some tweezers, particle manipulations are possible in Newtonian flows in the Stokes regime \cite{Petit2012}. Furthermore, the precise capture range and the critical translation speed of the manipulator have not yet been systematically investigated.

In this paper, we study a fluidic actuator based on a rotating nanowire. We build a mathematical model, find the exact velocity field for the flow generated by a tumbling spheroid in free space, numerically solve  the Stokes flow induced by a tumbling wire  that interacts  with a spherical particle, and examine the mechanism and condition in which the particle is captured and moving with the wire.   We also investigate the relations between the translation speed, geometric parameters, and the capture region.

\section{Model}\label{sect:model}

\begin{figure}
	\begin{center}
    \includegraphics[width=.50\textwidth]{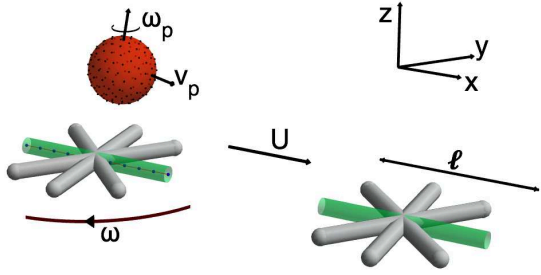}
  \caption{Schematic of the fluid tweezers. The  cylindrical nanowire with length $\ell$ and cross-section radius $r$ is rotating in a horizontal plane with an angular velocity $\omega \mathbf{e}_z$ while it also is translating with a constant velocity $U\mathbf{e}_x$. The sphere represents the  target particle with a radius $R$ in the flow.  The black dots along the nanowire's center-line and on the surface of the sphere illustrate the locations of the Stokeslets. }\label{fig:config}
  	\end{center}
\end{figure}

\subsection{Configuration of the problem} 
The fluidic tweezers are made of a magnetic nanorod or nanowire controlled by  magnetic fields. In our numerical model, the nanowire is modeled as a long cylinder with a length $\ell$ and a radius $r$. The target particle is a rigid sphere with radius $R$ in the fluid (illustrated in  figure~\ref{fig:config}). The wire is translating with a uniform velocity $U \mathbf{e}_x$ and rotating with an angular velocity $\omega\,\mathbf{e}_ z$. Without loss of generality, the wire is initially placed horizontally in the $x-y$ plane. The position of the centerline of the wire is prescribed as
\begin{eqnarray}
 \mathbf{x}=R_{\omega} s \, \mathbf{e}_ x
+  U\, t \,\mathbf{e}_ x,   
\label{rod_centerline}
\end{eqnarray}
where 
\begin{eqnarray}
R_{\omega}=\left(\begin{array}{crc}
\cos(\omega t)& -\sin(\omega t)& 0 \\
 \sin(\omega t)&\cos(\omega t)& 0 \\
 0 & 0 & 1 \\
\end{array}\right)
\end{eqnarray}
is the rotation transformation matrix ,  and $-\ell/2 \leq s \leq  \ell/2$.  The velocity of  the rod can be computed as 
\begin{eqnarray}
\mathbf{U}_{w}
=\omega  s
\left[
 -\sin(\omega t) \,\mathbf{e}_ x +
\cos(\omega t) \,\mathbf{e}_ y \right]+ 
U  \mathbf{e}_ x.
 \label{rod_vel}
\end{eqnarray}

We consider the particle as neutrally buoyant in the fluid and that Brownian motion is negligible. Assuming the particle is centered at $\mathbf{x}_p$, the translation velocity at the center is $\mathbf{v}_p$, and  the angular velocity about its center is $\boldsymbol{\omega}_p$, the surface velocity $\mathbf{U}_{p}$ can be computed as
\begin{equation}\label{protate}
\mathbf{U}_{p}= \mathbf{v}_p+(\mathbf{x}_0-\mathbf{x}_p) \times \boldsymbol{\omega}_p,
\end{equation}
for any $\mathbf{x}_0$ on the surface of the particle $\partial \Omega_p$.

For flows induced by micro-fluid tweezers, the characteristic Reynolds number is $\mbox{\textit{Re}}=U\,\ell\, \rho/\mu$, based on the translation speed, or $\mbox{\textit{Re}}=\rho\, \omega \,\ell^2/\mu$, based on the rotational speed. For tweezers and particle size below 100$\mu$m, $Re$ is usually much less than one and the inertial effect is negligible. Therefore, we assume the flow is in the Stokes regime. Also,  we assume that the fluid is incompressible. Then, the governing equations are reduced from the Navier-Stokes equations to the Stokes equations
\begin{eqnarray}\label{Stokes}
\mu \nabla^2 \mathbf{u}-\nabla p+ \mathbf{f} = 0, \quad
\nabla\cdot \mathbf{u}=0,
\end{eqnarray}
where $\mathbf{u}$ is the fluid velocity,  $p$ denotes the fluid pressure, $\mathbf{f} $ is  the external force exerted on the fluid, and $\mu$ is the dynamic viscosity.  

The flow is at rest without the wire and therefore the boundary conditions satisfy
\begin{eqnarray}\label{BD}
\begin{array}{rcl}
\mathbf{u}(\mathbf{x}_0)&=&\mathbf{U}_{w,p}(\mathbf{x}_0),\\
\displaystyle{\lim_{|\mathbf{x}|\to\infty}}\mathbf{u}(\mathbf{x})&=&0,
\end{array}
\end{eqnarray}  
where $\mathbf{x}_0$ is on the boundary of the wire $\partial \Omega_w$ or the boundary of the particle $\partial \Omega_p$. 

\subsection{Exact solution for flow induced by a tumbling spheroid nanowire}
For a tumbling rod with a prolate spheroid shape, the exact solution for the induced flow can be constructed from fundamental singularity solutions \cite{Blake1971, Pozrikidis1997, Zhao2010, Camassa2011JFM}.  Here we consider two reference frames. In the lab frame $\mathbf{x}$, the rod is tumbling with uniform translating velocity $U \mathbf{e}_x$ and rotating with angular velocity  $\omega\mathbf{e}_ z$. 
The centerline of the rod  is prescribed as \eqref{rod_centerline} and its velocity is \eqref{rod_vel}.
In the body frame $\mathbf{x}_b$, the rod is fixed. Without loss of generality, we assume the body is along the $x$-axis.
Then, the centerline of the rod is $(x_b,y_b,z_b)=(s,0,0)$, where $-\ell/2\leq s\leq \ell/2$. 

The relation between the body frame $\mathbf{x}_b$ and the lab frame $\mathbf{x}$ is 
\begin{eqnarray} \label{body-lab-vel}
\mathbf{x}=R_{\omega}\mathbf{x}_b+ U\, t\, \mathbf{e}_x,  \mbox{ or }  \mathbf{x}_b=R_{\omega}^{T}
\left(
\mathbf{x}-U\, t\, \mathbf{e}_x
\right).
\end{eqnarray}
The relation of the body frame velocity $\mathbf{v}_b$ and the lab frame velocity $\mathbf{u}$ is
\begin{eqnarray*}
\mathbf{v}_b
=
R_{\omega}^T
\left(
\mathbf{u}-
U  \mathbf{e}_x
\right)
+\omega \left(\begin{array}{ccc}
0& 1& 0 \\
 -1&0& 0 \\
 0 & 0 & 0 \\
\end{array}\right)
\mathbf{x}_b.
\end{eqnarray*}


Since $\mathbf{u}(\mathbf{x})=0$ as $|\mathbf{x}|\to\infty$, the equation above gives the background flow in the body frame:
\begin{eqnarray}\label{backgroudV}
\mathbf{v}_b(\mathbf{x}_b)&=&
U\left(- \cos(\omega t+\phi)\mathbf{e}_x +\sin(\omega t+\phi)  \mathbf{e}_y\right)
\nonumber\\
&&+\omega \left(y_b \mathbf{e}_x -x_b\mathbf{e}_y \right),
\end{eqnarray}
which is the combination of a uniform flow and a rotation flow. The boundary conditions become
\begin{eqnarray}
\mathbf{v}(\mathbf{x}_{b,0})&=&0,\\
\lim_{|\mathbf{x}_b|\to\infty}\mathbf{v}(\mathbf{x})&=&\mathbf{v}_b(\mathbf{x}_b),
\end{eqnarray}
where $\mathbf{x}_{b,0}$ is the on the surface of the rod in the body frame.

The surface of the spheroid rod is described by
\begin{equation}
\frac{x^2}{a^2}+\frac{y^2+z^2}{b^2}=1,
\end{equation}
where $a$ is the major semi-major axis and $b$ is the semi-minor axes of the spheroid.  The related focal length $2c$ and the eccentricity $e$ are
\begin{equation}
c=\sqrt{{a^2}-{b^2}}= e a.
\end{equation}
Based on the results of Chwang \& Wu \cite{Wu1975}, we  employ a line distribution of singularities between the foci $x=-c$ and $c$.  The detailed derivation and a  list of  singularities used for this study are provided in the Appendix.
 For a background flow \eqref{backgroudV} past a spheroid rod, the velocity field is 
\begin{eqnarray}\label{exact_vel}
&&\mathbf{u}(\mathbf{x})=U_1\mathbf{e}_x+U_2 \mathbf{e}_y+\omega  y \mathbf{e}_x-\omega  x \mathbf{e}_y 
-\alpha _1\int _{-c}^c\mathbf{u}_{\rm S}\left(\mathbf{x}-\bdxi ;\mathbf{e}_x\right)d\xi
-\alpha _2\int _{-c}^c\mathbf{u}_{\rm S}\left(\mathbf{x}-\bdxi ,\mathbf{e}_y\right)d\xi \nonumber \\
&&+\beta _1\int _{-c}^c\left(c^2-\xi ^2\right)\mathbf{u}_{\rm D}\left(\mathbf{x}-\bdxi ;\mathbf{e}_x\right)d\xi 
+\beta _2\int _{-c}^c\left(c^2-\xi ^2\right)\mathbf{u}_{\rm D}\left(\mathbf{x}-\bdxi ,\mathbf{e}_y\right)d\xi
\\
&&+\chi _1\int _{-c}^c\left(c^2-\xi ^2\right)\mathbf{u}_{\rm{SS}}(\mathbf{x}-\bdxi ;\mathbf{e}_x,\mathbf{e}_y)d\xi 
+\chi _2\int _{-c}^c\left(c^2-\xi ^2\right)^2\frac{\partial }{\partial  y} \mathbf{u}_{\rm D}\left(\mathbf{x}-\bdxi ;\mathbf{e}_x\right)d\xi 
\nonumber \\&&
+\chi _3\int _{-c}^c\left(c^2-\xi ^2\right)\mathbf{u}_{\rm R}\left(\mathbf{x}-\bdxi ;\mathbf{e}_z\right)d\xi ,\nonumber 
\end{eqnarray}
where
\begin{eqnarray*}
U_1= - U \cos\left(\omega t+\phi\right),&\quad&
U_2= U\sin\left(\omega t+\phi\right),\\
\chi _1=\frac{e^2 \omega }{-2 e+\left(1+e^2\right) L_e},&\quad&
\chi _2=\frac{\omega -e^2 \omega }{-8 e+4 \left(1+e^2\right) L_e},\quad
\chi _3=\omega \frac{2-e^2}{-2 e+\left(1+e^2\right)L_e},
\end{eqnarray*}
and 
\begin{eqnarray*} 
L_e&=&\log \left(\frac{1+e}{1-e}\right).
\end{eqnarray*}
The singularities are on the location $\boldsymbol{\xi}=\xi\mathbf{e}_x$.

All integrals in the velocity field \eqref{exact_vel} can be integrated explicitly and we can use the relation \eqref{body-lab-vel} to convert velocities between the body frame and the lab frame. In other words, we have found the explicit expression of the velocity field.

\subsection{Numerical method for the fluidic tweezers and coupled system}
To solve the nanowire coupled with non-zero volume particles, we implement the regularized Stokeslet method developed by Cortez \cite{Cortez2001}. 
 The idea of regularized Stokeslets is to replace the singular force for Stokeslet with a small, smooth, and concentrated force ($\mathbf{f}=\mathbf{f}_0\phi_\epsilon(\mathbf{x}-\mathbf{x}_o)$) so that the velocity is defined everywhere, including the location of the force $\mathbf{x}_o$. 
The regularization parameter $\epsilon$  approximately represents the radius of the blob.  When $\epsilon \to 0$, 
$\phi_\epsilon(\mathbf{x}-\mathbf{x}_o)\to \delta(\mathbf{x}-\mathbf{x}_o)$.
We choose the regularized Stokeslet given by Cortez  \cite{Cortez2001} and denote the regularized Green's function as $\mathbf{G}(\mathbf{x},\mathbf{x}_o,\epsilon)$.  The blob is 
\begin{equation*}
\phi_\epsilon(\mathbf{x})=\frac{15\epsilon^4}{8\pi ({|\mathbf{x}|}^2+\epsilon^2)^{7/2}}.
\end{equation*}
The velocity of the flow with a single regularized Stokeslet at $\mathbf{x}_o$ is
\begin{eqnarray*}
&& \mathbf{U}_{\rm s}(\mathbf{x};\mathbf{f}_o)=\mathbf{G}(\mathbf{x},\mathbf{x}_o,\epsilon)\cdot \mathbf{f}_o
  \\
 &&=\frac{1}{8\pi}\left[
 \frac{\mathbf{f}_o}{(|\mathbf{x}-\mathbf{x}_o|^2+\epsilon^2)^{1/2}}+\frac{\epsilon^2\mathbf{f}_o}{(|\mathbf{x}-\mathbf{x}_o|^2+\epsilon^2)^{3/2}}
 + \frac{(\mathbf{f}_o\cdot(\mathbf{x}-\mathbf{x}_o))(\mathbf{x}-\mathbf{x}_o)}{(|\mathbf{x}-\mathbf{x}_o|^2+\epsilon^2)^{3/2}} \right].
\end{eqnarray*}
More details of these formulae can been found in previous references~\cite{Cortez2001,Buchmann2015,Zhao2015}.

To solve the problem with the proper boundary conditions, we distribute the regularized Stokeslets evenly along the center-line of the wire (see figure \ref{fig:config}).  The regularization parameter on the centerline of the wire, $\epsilon_w$, is chosen as the radius of the wire $r$.
For the particle, regularized Stokeslets are distributed evenly on the surface of the particle and locations of the Stokeslets are obtained by using Spherical Centroidal Voronoi Tessellation (SCVT) and the package STRIPACK \cite{Renka1997,Du2003}. 


At any instantaneous time, the strength $\mathbf{F}(\mathbf{x})$ of a regularized Stokeslet is determined by the no-slip boundary conditions on the wire and the spherical particle. The total velocity induced by all the regularized Stokeslets in the flow is given by
\begin{eqnarray}
\mathbf{u}(\mathbf{x},t) &=& \sum_{m=1}^{N_w}  \mathbf{G}(\mathbf{x}, \mathbf{x}_m;\epsilon_w) \cdot \mathbf{F}(\mathbf{x}_m)+
 \sum_{k=1}^{N_p}  \mathbf{G}(\mathbf{x}, \mathbf{x}_k;\epsilon_p) \cdot \mathbf{F}(\mathbf{x}_k) ,
 \label{u_F}
\end{eqnarray}
where $\mathbf{x}_m$ is the position of the $m$th regularized Stokeslet on the wire,  $\mathbf{x}_k$ is the position of the $k$th regularized Stokeslet on the particle, and $\epsilon_p$ is the regularization parameter on the particle.

Without the finite-sized particle embedded in the flow, the boundary conditions \eqref{BD} are applied on the wire only and  yield 
\begin{equation}
\sum_{m=1}^{N_w}  \mathbf{G}(\mathbf{x}_i, \mathbf{x}_m;\epsilon_w) \cdot \mathbf{F}(\mathbf{x}_m) =\mathbf{U}_w(\mathbf{x}_i),
\label{Fonwire}
\end{equation}
where  $\mathbf{x}_i\in \partial \Omega_w$, $i=1,\cdots, N_w$. 
These equations can be rewritten in matrix form with $3N_w$ unknown variables. The linear system is then solved numerically 
to obtain the regularized Stokeslet strengths $\mathbf{F}_m=\mathbf{F}(\mathbf{x}_m(t))$.
 Consequently, the fluid velocity field $\mathbf{u}(\mathbf{x},t)$ can be reconstructed everywhere using \eqref{u_F}. 

When a finite-sized particle presents in the system, besides the boundary condition on the wire  
\begin{eqnarray}
 &&\sum_{m=1}^{N_w}  \mathbf{G}(\mathbf{x}_i, \mathbf{x}_m;\epsilon_w) \cdot \mathbf{F}(\mathbf{x}_m)+
 \sum_{k=1}^{N_p}  \mathbf{G}(\mathbf{x}_i, \mathbf{x}_k;\epsilon_p) \cdot \mathbf{F}(\mathbf{x}_k)
=\mathbf{U}_w(\mathbf{x}_i), i=1,2,\cdots, N_w,
\label{Fonwire2}
\end{eqnarray}
additional boundary conditions $\mathbf{u}(\mathbf{x}_k) = \mathbf{U}_p(\mathbf{x}_k)$ ($k=1,2,\cdots,N_p$) should also be satisfied. However, the velocities on the particle $\mathbf{U}_p(\mathbf{x}_k)$ are unknowns. By substituting $\mathbf{U}_p$ with equation \eqref{protate}, we obtain relations
\begin{eqnarray}
&&\sum_{m=1}^{N_w}  \mathbf{G}(\mathbf{x}_j, \mathbf{x}_m;\epsilon_w) \cdot \mathbf{F}(\mathbf{x}_m)+ \sum_{k=1}^{N_p}  \mathbf{G}(\mathbf{x}_j, \mathbf{x}_k;\epsilon_p) \cdot \mathbf{F}(\mathbf{x}_k) 
-\mathbf{v}_p-(\mathbf{x}_0-\mathbf{x}_p) \times \boldsymbol{\omega}_p=0,
 \label{Fonsphere}
\end{eqnarray}
where $\mathbf{x}_i\in \partial \Omega_p$, $i=1,\cdots, N_p$. 
The unknown variables in equations \eqref{Fonwire2} and  \eqref{Fonsphere} are $\mathbf{F}$, $\mathbf{v}_p$, and $\boldsymbol{\omega}_p$, which add up to $3(N_w+N_p)+6$ unknowns. Since equations \eqref{Fonwire2} and \eqref{Fonsphere} correspond to only $3(N_w+N_p)$ scalar equations, 6 additional equations are required to determine the system. They can be derived from the non-inertia approximation of the sphere. Because the inertia is negligible, the total force and total torque on the particle are both zero.  These imply
\begin{eqnarray}\label{FTzero}
\sum_{k=1}^{N_p} \mathbf{F}(\mathbf{x}_k)&=&0 ,  \\
\sum_{k=1}^{N_p} (\mathbf{x}_k-\mathbf{x}_p) \times \mathbf{F}(\mathbf{x}_k)&=&0.
\end{eqnarray}
After the force strength $\mathbf{F}(\mathbf{x})$ and the rigid body motion for the spherical particle $\mathbf{v}_p$ and $\boldsymbol{\omega}_p$ are determined, the velocity of the flow can be evaluated with the equation \eqref{u_F}.

\begin{figure*}[tbh!]
	\begin{center}
    \includegraphics[width=.8\textwidth]{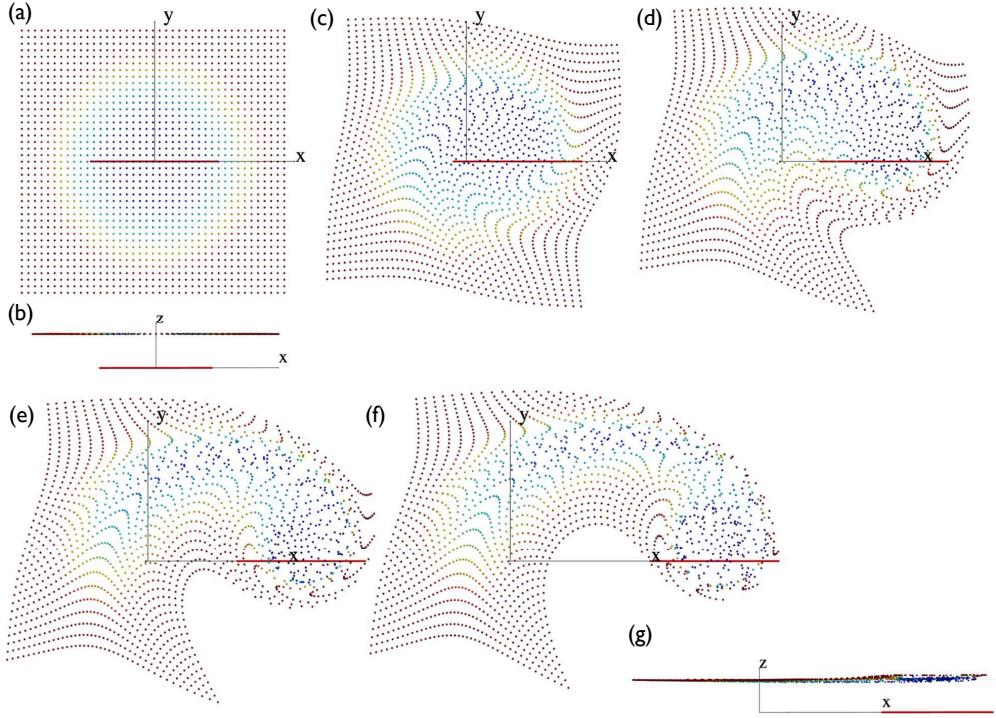}
   \caption{Snapshots of fluid tracers in the Stokes flow generated by the tumbling wire. The translation speed of the wire is $U=0.08$. These fluid tracers are initialized on a flat plane above the wire at $z=0.3$.  The fluid tracers are colored according to their initial distance from the $z$-axis. (a) and (b): top view and side view of the initial setup; (c)-(e): top views at $t=5$, $10$, and $15$, respectively; (f) and (g): top view and side view at $t=20$. The red thick lines represent the centerline of the wire. (Multimedia view)}
   \label{fig:fluidtracers}
  	\end{center}
\end{figure*}

\subsubsection*{Nondimensionalization, parameters and errors}
In our simulations,  we use dimensionless parameters so that the length of the wire $\ell =1 $, rotation rate $\omega= -2\pi$, and the period of the rotation $T=1$. As such, the value of $U$ actually represents the relative time scale between rotation and translation. In order to compare with the results of the experiment in  Petit {\it et al.}  \cite{Petit2012}, the radius of the wire is set to 0.0067 and the radius of the sphere is set to 0.2. 
The numbers of regularized Stokeslet on the particle and on the wire are
$N_p=150$  and  $N_w =94$, respectively. 
Since the wire and its motion have reflection symmetry in the $z=0$ plane, the behavior of a particle below the tweezers is the same as those in its image position above the tweezers. Therefore, we only show the results for the particles or tracers initialized above the wire. 

 Since the boundary conditions are enforced only at those discretized points where Stokeslets are centered, relatively larger errors may occur between the Stokeslet points. Increasing the regularization parameter can decrease such errors, but might introduce more errors in the computed forces \cite{Cortez2001}. The regularization parameter on the surface of particle is chosen as 0.4 times the estimated average distance between two points, that is $\epsilon_p=0.4\sqrt{4\pi R^2/N_p}$, where the factor 0.4 is a numerical optimal value for the discretized spherical particle \cite{cortez2005method, Martindale2016}.  For a sphere at typical positions in the region ($-0.5<x<0.5$, $-0.5<y<0.5$, $0.25<z<0.6$), the velocity error on a random point on the sphere is about 4\% of the average magnitude of the velocity on the sphere. A similar test shows that the velocity error along the wire's center-line is less than 3\%. To test whether the velocity of the sphere can be sufficiently described by the Stokeslets, we increased the number of the points on the sphere to 400 and found only 0.2\% difference from the 150-point case. 



\section{Results}

With our model, we examine the flow induced by the fluid tweezers, focus on the spherical particle captured by the fluid tweezers, and attempt to explain the capture mechanism.  

Before we couple non-zero volume particles with the tweezers in the system,  we  examine the fluid tracers in the flow induced by the wire, for which the exact velocity field is obtained. The movement of the fluid tracers in the flow is shown in figure \ref{fig:fluidtracers}  (Multimedia view). 
We find that some particles starting in the vicinity of the wire follow the motion of the wire and some fall behind. Little motion is observed in the vertical ($z$) direction.  

\begin{figure}[b!]
	\begin{center}
    	\includegraphics[width=.5\textwidth]{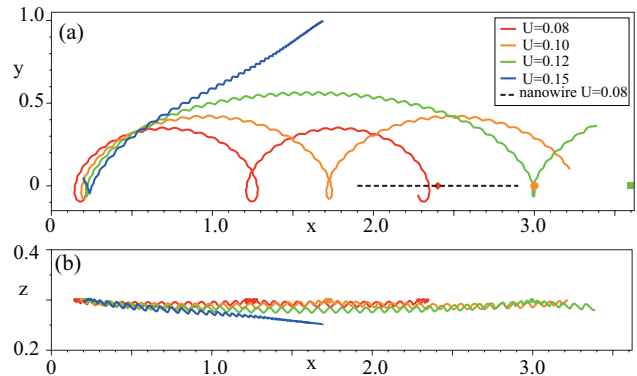}
  \caption{Top view (a) and side view (b) of the trajectories of a spherical particle in the Stokes flow generated by a tumbling  wire at different translation speeds $U=0.08, 0.10, 0.12$ and $0.15$. The center of the sphere is initialized  at (0.2, 0.05, 0.3) and its radius is $R=0.2$. The total time for the simulations is 30.   The dashed line indicates the position of the wire with the smallest  $U=0.08$ at $t=30$.  The solid markers (red diamond, orange dot, and green square) represent the center of the wire at $t=30$ for corresponding trajectories. The consistent color scheme is used to represent different velocities for the trajectories and the centers of the wire. The center of the wire for the largest translation speed $U=0.15$ is outside the plotting window.  Different scales are adopted for the vertical axes in (a) and (b) to make the trajectories observable. (Multimedia view)}
\label{fig:traj}
  	\end{center}
\end{figure}

By examining the trajectories of individual spherical particles, we find that their trajectories are similar to those of the tracer particles (figure~\ref{fig:traj}, multimedia view). The capture depends on the initial position of the particle and also on the translation speed. Note that since we normalize the angular velocity of the rotation of the wire, the effect of the wire's rotation speed is represented by the effect of the translation speed. The typical trajectory of a captured particle exhibits a trochoid-like curve with small wiggles, as the red, orange and green curves show in figure~\ref{fig:traj} (Multimedia view). The shape of the curves depends on the particle's initial position and the translation speed of the wire. When the translation speed is large, the particle may deviate from the prescribed route of the wire. The trajectories of the trapped particles are almost planar, while the untrapped particle moves along a trajectory (the blue curve in figure~\ref{fig:traj}b) with a relatively large movement  in the $z$ direction, perpendicular to the plane of the motion of the wire. 



\begin{figure}[t!]
	\begin{center}
    \includegraphics[width=.450\textwidth]{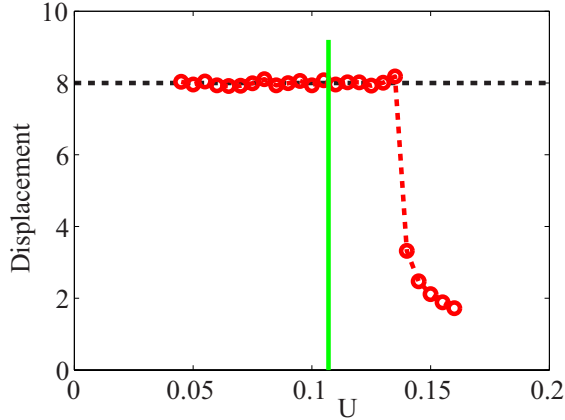}
  \caption{The displacement of the particle in the $x$ direction (red circles) as a function of the translation speed of the wire when the center of the wire reaches $x=8$. The particle is initially placed at (0,\,0,\,0.3). The dashed red line is used to guide the eyes, the black dotted line indicates the displacement of the wire, and the green line indicates the speed at which capture has been achieved experimentally by Petit {\it et al.} \cite{Petit2012}. }\label{fig:capture_speed}
  	\end{center}
\end{figure}

To distinguish the successful capture cases of a particle from the no-capture cases and quantify the capture region, we measure the horizontal movement of the spheres or tracer particles when the wire translates over a specified distance. We choose a total distance of 8, which corresponds to a total time $t=8/U$. If the displacement of the particle in the forward ($x+$) direction relative to the rod center is less than 1, we consider the particle has been captured (figure~\ref{fig:capture_speed}). The results with a larger displacement are qualitatively the same and will be discussed in the Discussion section. As shown in figure~\ref{fig:capture_speed}, a particle with a radius of 0.2 and starting 0.3 above the center of the wire is captured when the translation speed is below about 0.13. This is consistent with the experiment reported in the previous study by Petit {\it et al.}\cite{Petit2012} that a 6$\mu$m diameter particle can be captured by a 15$\mu$m long wire when the wire is rotating at 54\,Hz and translating at as high as 87\,$\mu$m/s, which corresponds to a lower limit of nondimensionalized maximal capture speed 0.11. The speed is marked as the green line in figure 4.

\begin{figure*}
	\begin{center}
    \includegraphics[width=1.0\textwidth]{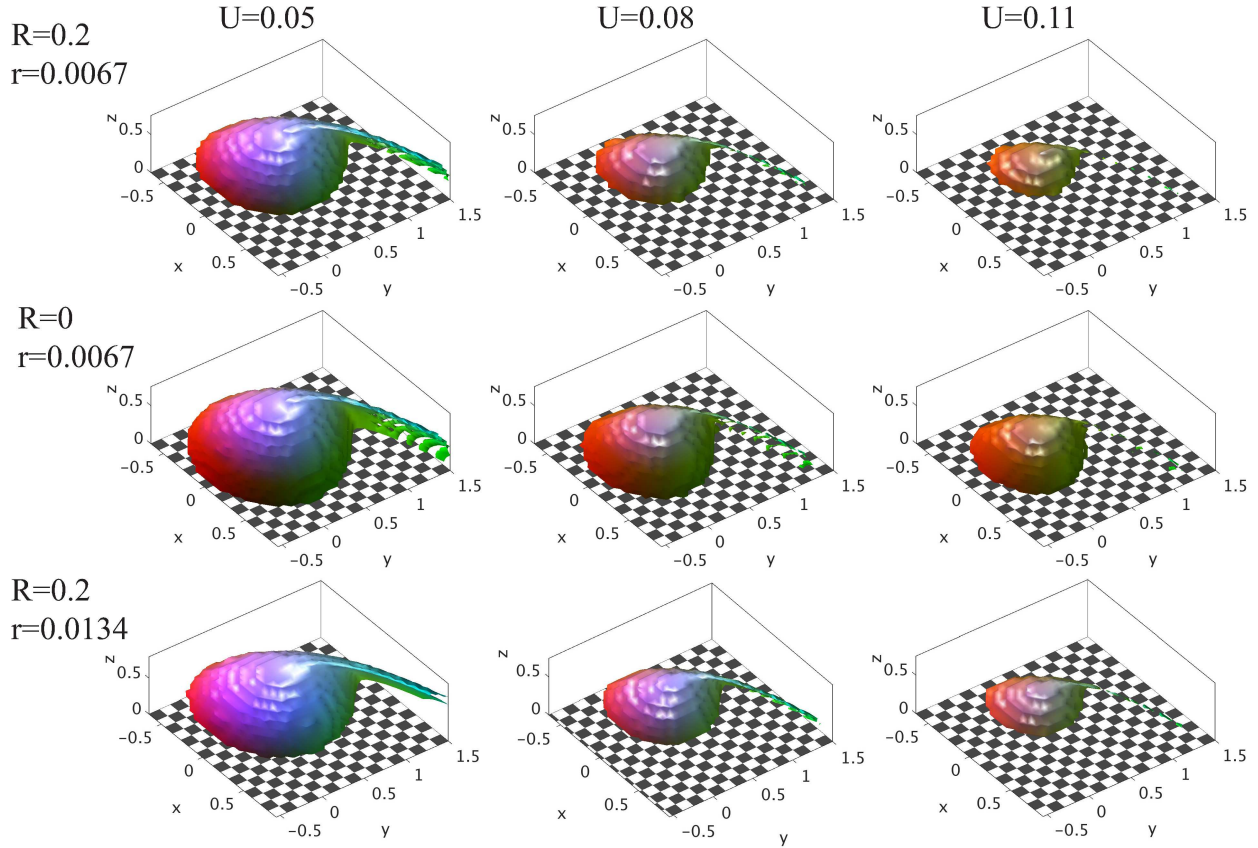}
  \caption{Capture regions of different dimensions of the wire and particle (rows) and the translation speeds (columns) of the wire. The boundaries of the capture region are shown above $z=0.25$ as colored surface for a rigid sphere and the boundaries above $z=0.05$  are shown for a tracer particle. The color (i.e. (R,G,B) value) on the surface depends on the local position (i.e. (x,y,z)).
  }\label{fig:captureregion}
  	\end{center}
\end{figure*}

To quantify the condition in which the particles are captured, we systematically vary the initial condition of the spherical particle for three translation speeds of the wire and two particle sizes. Since the flow  is symmetric, we only examine the initial position in the $z+$ half space, i.e., above the  plane where the wire is, for  figure 5. For the tracer particle ($R=0$), the initial positions are considered in a box $(-0.65\leq x \leq 0.9, -0.55\leq y \leq 1.5,0.05 \leq z \leq 0.75)$, with a grid resolution of 0.05 in all three dimensions. For the particle with a radius of 0.2, the box with $z\geq 0.25$ is considered to avoid overlapping between the wire and particles at the initial state. 
The function {\it isosurface} in MATLAB (ver. 8.0.0.783, Mathworks, Inc.) is used to find the isosurface with a horizontal movement between 7 and 9 in the $x+$ direction. Then, inside the surface is the capture region. The space in figure~\ref{fig:captureregion} is color coded so that the position of the surfaces can be compared point by point.

We find that overall the capture regions have pear-like shapes, as shown in figure~\ref{fig:captureregion}. The cross sections in the $x$-$y$ plane are pointier in the $y+$ side. The green pointy region connects to a thin and scattered region, which phases out between the $y+$ and $x+$ direction. The capture region shrinks as the translation speed $U$ increases. In the region both accessible by the tracer particles and spherical particles with radius $R=0.2$, the capture regions are nearly identical. Because a larger range of $z$ is used for the tracer particle, the capture region for the tracer particle in the second row may look bigger than that for the sphere in row 1 in figure~\ref{fig:captureregion}. 

Besides comparing the capture region for the passive tracer particle,  we also examined the capture region of spheres by a wire with the wire's radius doubled (i.e., $r=0.0134$). From row 1 and row 3 in figure~\ref{fig:captureregion}, we find that the capture region expands in all three dimensions slightly.

\begin{figure*}
	\begin{center}
    \includegraphics[width=1.0\textwidth]{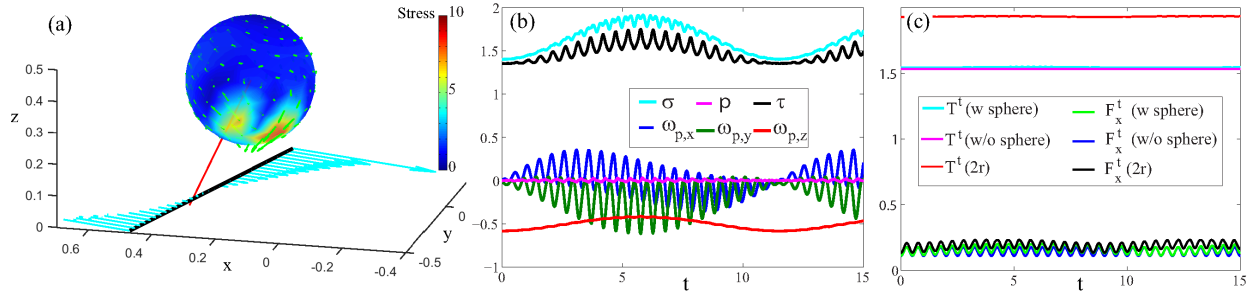}
  \caption{Stresses and forces on the sphere and wire. (a) The force on the sphere and the wire exerted by the fluid at an example time $t=4.2$. The color on the sphere represents the magnitude of the total stress. The red arrow represents the rotation vector $\omega_p$. (b) The average magnitude of the stress $\sigma$, the normal component $p$ of the stress, the parallel component $\tau$ of the stress, and the three components of the rotation of the sphere as a function of time. $r=0.0067$, $U=0.08$. (c) The magnitude of the total torque $T^t$ and the total force in the $x$ direction $F^t_x$ as a function of time for different configurations. ``2r'' refers to the case where $r=0.0134$. $R=0.2$ and $U=0.08$. 
  }\label{fig:force}
  
  	\end{center}
\end{figure*}

To further explore the influence of the capture on the particle and wire, we examine the rotation and the stress of the sphere and the force distribution on the tweezers (figure~\ref{fig:force}). We find that the rotation of the sphere about its center is mainly in the $-z$ direction, which is the same as the rotation of the wire. There are two frequencies in the rotation (figure~\ref{fig:force}b\,\&\,c). The higher frequency is twice  the frequency of rotation of the wire. This high frequency is due to the periodic motion of the two  ends of the wire relative to the sphere. The lower frequency corresponds to the periodic motion of the sphere relative to the center of the wire. For the stress $\sigma$ on the sphere, we decompose it into the normal component $p$, which is normal to the surface, and the shear stress $\tau$, which is parallel to the surface. The normal stress $p$ is negligible compared to the shear street $\tau$. Taking the average over the surface of the sphere, the magnitude of the shear stress is about 1.5 in dimensionless units. The flow exerts the largest shear stress on the sphere near the close encounter point for the wire and the sphere.

Even though the wire and sphere are coupled in the system, the total torque $T^t$ and net force $F^t$ on the wire are nearly independent of the presence of the sphere (figure \ref{fig:force}c). The magnitude of the torque is about 1.5 in dimensionless units and is about $3\times10^{-18}$N$\cdot$m using experimental parameters. 
The component of the total force in the translation direction ($\text{F}_x^t$) is about 0.15 in dimensionless unit or $2.5\times10^{-14}$N in physical units. For a wire with a doubled diameter, the torque and net force on the wire are both increased by about 25\%. Such an increase is consistent with the slender body theory that the force on a slender rod is proportional to the slenderness parameter $\mathrm{ln}^{-1}(\ell/r)$ with a fixed velocity \cite{Batchelor1970}.

\section{Discussion}

\begin{figure*}
	\begin{center}
    \includegraphics[width=.90\textwidth]{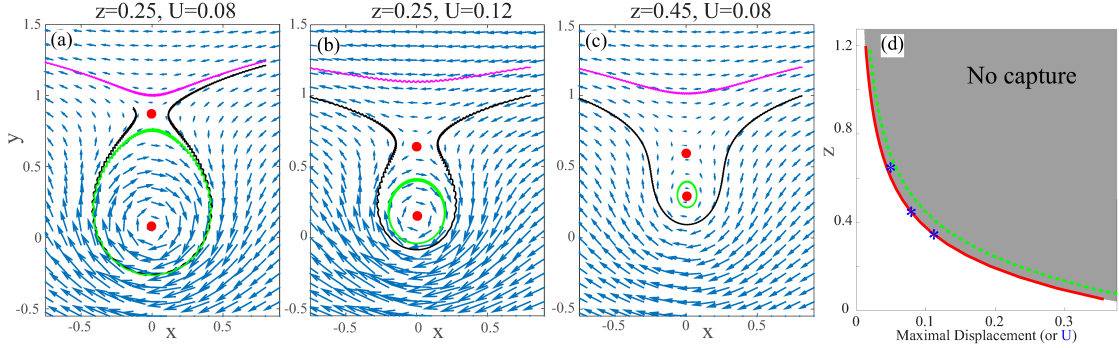}
  \caption{Displacement field analysis for tracer particles. (a)-(c) are displacement fields 
   in the horizontal plane, in a moving frame translating at the same speed of the center of the wire.
(a) and (b) are at $z=0.25$ with the translation speed $U=0.08$ and $U=0.12$, respectively.   (c) is with $U=0.08$ and at $z=0.45$, a different height to (a). Blue arrows represent the displacement. The red dots indicate the stagnation points. The magenta, green and black trajectories in (a)-(c) are  three representative tracer particles trajectories over a time period of 100. (d) is the capture diagram predicted by displacement field analysis. The red curve is the maximal displacement in the $x$ direction generated by a rotating wire with $r=0.0067$ at different heights $z$. Bounded by the red curve, the shaded region is the predicted region at which height $z$ and translating speed $U$ no capture occurs. The green dotted curve represents the maximal displacement to height relation for a thicker wire with $r=0.0134$. The blue asterisks mark the observed maximal $z$ coordinate of the capture regions and the respective $U$ in the first row in figure \ref{fig:captureregion}.} \label{fig:displacement}
  	\end{center}
\end{figure*}

To reveal the capture mechanism and understand the capture region, we examine the displacement for tracer particles in the moving frame within
 one period of rotation of the wire (i.e., the vector field $\mathbf{d}=\mathbf{x}_p(t=1)-\mathbf{x}_p(t=0)$ in the moving frame). Since the capture is insensitive to the particle size, we consider the displacement of a tracer particle in the body frame translating with the center of the wire. The displacement field can be approximated as the combination of the translation displacement and a vortex. When the particle is far from the wire and the influence of the wire's rotation on the particle is weak, the displacement field is simply the uniform displacement $-UT$. A rotating wire generates a vortex in the fluid with maximal linear velocity of the fluid near the end points and the velocity decreases with increasing $z$. Note that the vortex has some detailed but complex features: (1) Due to the non-slip boundary condition, the fluid near the surface of the wire always moves with the wire and therefore the displacement is nearly zero near the surface. (2) The displacement field is not exactly axis-symmetric, since the initial orientation of the wire defines a direction. Since the motion of the wire is in the $x$-$y$ plane and the diameter of the wire is very small compared to the length, the displacement field is also nearly planar. The displacement in the $z$ direction is negligible ($< 0.007$ in the $z=0.25$ plane) in determining the overall capture region. Therefore, we focus on the displacement field in the $x$-$y$ plane at different heights $z$.

When the translation speed $U$ is small compared to the linear velocity of the wire and $z$ is small, the displacement field has two stagnation points along the $y+$ axis. As the linear shear velocity generated by the rotation of the wire increases from zero from the center of the wire and then decreases to zero as the distance goes to infinity, the two stagnation points appear where the forward ($x+$) velocity is canceled out by the uniform backward displacement.  Since the only region with forward velocity is between these two stagnation points, the stagnation points indicate the size and the position of the capture region.  The stagnation point with a smaller $y$ value is a stable stagnation point. A particle starting between these two stagnation points encircles this stable stagnation point with an oval shape (green curves in figure~\ref{fig:displacement}(a-c)). The stagnation point with a greater $y$ value is an unstable saddle point. A particle that starts from the stable manifold in the upstream $(x+,y+)$ takes an infinite time to reach the saddle point for a finite distance. Thereafter, the particle appears to be captured. Because the trajectory of a point near the stable manifold is sensitively dependent on the initial position, the apparent capture is sensitive to initial positions and numerical errors. Therefore, the particle captured is scattered in this region.  When the starting point is sufficiently far from the saddle point in the $y+$ direction, the uniform translation dominates and no capture can occur (the magenta curves in figure~\ref{fig:displacement} (a-c)). A similar simplified 2D kinematic model of atmospheric or oceanic flow was analyzed and similar capture regions were found by Flierl (1981)\cite{flierl1981particle}.

\begin{figure}
	\begin{center}
    \includegraphics[width=.50\textwidth]{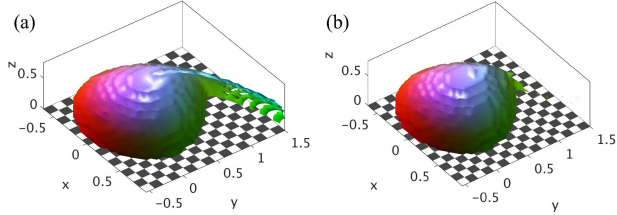}
  \caption{Comparison of the capture regions when the wire travels a distance of 8 (a) and 16 (b). $R=0$, $r=0.0067$, and $U=0.05$.}\label{fig:longdistance}
  	\end{center}
\end{figure}

As shown from the analysis, the particles in the scattered region near the stable manifold of the saddle point are not truly captured. They gradually fall behind the rod as the system evolves for a longer time and such spurious capture regions shrink.  Figure~\ref{fig:longdistance} shows a comparison of the capture regions when the rod travels with different distances, i.e., with a different total time for the simulations.

When $U$ or $z$ increases, the two stagnation points move closer, and eventually merge and disappear (figure \ref{fig:displacement}). Since the distance between the two stagnation points roughly determines the radius of the capture region in a $z$ cross sectional plane, the decrease of the distance as $z$ increases or $U$ increases explains the drop-like shape of the capture region and the loss of volume as $U$ increases. Essentially, canceling the displacement in the forward direction $\mathbf{d}_x$ in the vortex by $-UT$ determines the existence of the capture region. We examine the maximal displacement  in the $x$ direction in cross-sections of the vortex as a function of $z$ (figure \ref{fig:displacement}d). Such a plot indicates the height of the capture region at a given $U$ and outlines the capture region and the non-capture region. As shown in figure \ref{fig:displacement}c, the heights of the capture regions (asterisks) lie closely on this curve. With a thicker wire, the curve (green dotted) is shifted slightly upward, which is consistent with the larger capture regions by the thicker wire.


As the fluidic tweezers  often intend to gently manipulate biological cells, we compare the rough magnitudes of resultant stresses and forces on the spherical particle with the biological values. With the parameters in Petit  {\it et al.} \cite{Petit2012}, the forces and stresses on the sphere are in general gentle for biological cells. The magnitude of the stress on average is about 0.08\,Pa on the sphere. Such a magnitude of stress is sufficient to induce intracellular $\mathrm{Ca}^{2+}$ or $\mathrm{Ka}^{+}$ channel activation (0.02-0.4 Pa and 0.02-1.65 Pa, respectively), but unlikely to induce other biological response (typically $>$0.5 Pa) if the sphere is an endothelial cell \cite{davies1993mechanical}.  The stretching force is around 10\,pN; such a force is not sufficient to generate a large deformation of cells such as blood cells, who require a force in the magnitude of 100\,pN to deform significantly ($>$10\% diameter) \cite{dao2003mechanics}.

This study is limited to purely fluid mechanical effects of tweezers in free space. Since there are no other forces to overcome for the particle, the kinematic effects are dominant and thoroughly examined. Future studies might introduce other effects from real-life applications, such as forces from the contact between the particle and the surface, the gravity forces, buoyancy,  
boundaries of the fluid, and Brownian motion for particles with smaller sizes (typically $\le 1 \mu m $).
When the immersed particle is no longer a sphere, the trajectory of a particle may depend on both its shape and its initial orientation. Our preliminary study suggests that the capture regions are qualitatively similar  when the sphere is replaced by an oblate or prolate spheroid. However, the details of the capture region depends on the particle's shape and initial orientation in a non-trivial way. When the inertial force plays a role in the flow beyond the Stokes regime, the flow fields change qualitatively. Due to centrifugal forces, outward motions are expected as the sphere and fluid rotate. As such, the flow may become fully three-dimensional and the stagnation points in the planar displacement field are not sufficient to describe the capture region. Thorough investigations on these effects will be carried out  in the near future.
 
\section{Conclusions}

In conclusion, we have studied a tumbling nanowire that manipulates micro-objects.  For the spheroid tweezers, we derive an exact velocity field for the flow with the singularity method \cite{Wu1975}; in the presence of a  neutrally buoyant  spherical particle, we study the flow with the regularized Stokeslet method. For a  particle in a viscous/micro fluid, the kinematic effect of the vortex generated by the fluidic tweezers is responsible for the capture of the particles.
 The trajectories of the particles can be understood from the combination of a vortex and a uniform translation. The stagnation points in the combined displacement field provide key information on the shape and size of the capture regions. We found that the behavior of a finite-size sphere is nearly the same as a tracer particle for typical sizes reported in the experiments. Furthermore, the external forces and torques to drive the nanowire vary little for different particles.

\begin{acknowledgments}
Funding for Y.D. was provided by NSAF-NSFC grant No. U1530401, and the Recruitment Program of Global Young Experts.
\end{acknowledgments}

\appendix
\section{Exact solution for a spheroid tumbling in free space}

When the rod is  in the shape of a prolate spheroid, the exact solution for the Stokes flow generated by the spheroid is derived in this appendix. Firstly, we provide a  list of  singularities used for this study. The Stokeslet  $\mathbf{u}_{\rm S}$ is a fundamental solution of the Stokes equation for a single point force,
\begin{eqnarray}
\mu\nabla^2\mathbf{u}+\mathbf{f}_{\rm S}&=&\nabla p,\nonumber\\
\nabla\cdot \mathbf{u}&=&0, \label{stokes_app}
\end{eqnarray}
where  $\mathbf{f}_{\rm S}=8\pi \mu \bda \delta(\mathbf{x})$, $\bda $ is the strength of the singularity located at the origin,  and  $ \delta(\mathbf{x})$ is the 3D Dirac delta function at the origin. The explicit formulae of Stokeslet  are
\begin{eqnarray*}
 \mathbf{u}_{\rm S} (\mathbf{x};\bda)&=&\frac{\bda}{|\mathbf{x}|}
+\frac{(\mathbf{x} \cdot \bda )\mathbf{x} }{ |\mathbf{x}|^3}\\
p_{\rm S} (\mathbf{x};\bda)&=&-2\mu\frac{\mathbf{x} \cdot \bda}{|\mathbf{x}|^3},
\end{eqnarray*} 
where $\mathbf{x}=(x,y,z)$. 

Due to the linear property of the Stokes equations, a derivative to any order of $\mathbf{u}_{\rm S}$  and $p_{\rm S} (\mathbf{x};\bda)$ is also a solution of  \eqref{stokes_app} with corresponding derivative of $\mathbf{f}_{\rm S}$.
The Stokes doublet is
\begin{eqnarray*}
&& \mathbf{u}_{\rm{SD}}(\mathbf{x};\bda,\bdb)= -(\bdb\cdot \nabla)
   \mathbf{u}_{\rm S} (\mathbf{x};\bda) 
   =  \frac{(\bdb\times \bda)\times \mathbf{x}}{ |\mathbf{x}|^3 }-  \frac{(\bda.\bdb) \mathbf{x}}{ |\mathbf{x}|^3} + 3
   \frac{( \bda\cdot \mathbf{x}) (\bdb\cdot \mathbf{x})\mathbf{x}}{ |\mathbf{x}|^5} .
   \end{eqnarray*}
   The symmetric component of a  Stokes doublet is a stresslet
\begin{eqnarray*}
&&\mathbf{u}_{\rm SS}(\mathbf{x};\bda,\bdb) =\frac{1}{2}[
   \mathbf{u}_{\rm SD}(\mathbf{x};\bdb,\bda) +
    \mathbf{u}_{\rm SD}(\mathbf{x};\bda,\bdb)] 
    =  -\frac{(\bda\cdot\bdb)\mathbf{x}}{ |\mathbf{x}|^3} +\frac{3 (\bda\cdot \mathbf{x}) (\bdb\cdot \mathbf{x})\mathbf{x}}{ |\mathbf{x}|^5}.
   \end{eqnarray*}
The antisymmetric component of a Stokes doublet is Rotlet or couplet
\begin{eqnarray*}
&&\mathbf{u}_{\rm R}(\mathbf{x}; \boldsymbol{\gamma}) =
\frac{1}{2}[
   \mathbf{u}_{\rm SD}(\mathbf{x};\bdb,\bda) -  \mathbf{u}_{\rm SD}(\mathbf{x};\bda,\bdb)] 
    =\frac{1}{2} \nabla\times \mathbf{u}_{\rm S} (\mathbf{x}; \boldsymbol{\gamma}) = \frac{\boldsymbol{\gamma} \times \mathbf{x}}{|\mathbf{x}|^3},\end{eqnarray*}
    where $\boldsymbol{\gamma} =\bda\times \bdb$.
A potential doublet is
\begin{eqnarray*}
 \mathbf{u}_{\rm D}(\mathbf{x}; \boldsymbol{\delta}) =
\frac{1}{2} \nabla^2\times \mathbf{u}_{\rm S} (\mathbf{x}; \boldsymbol{\delta})
 = -\frac{\boldsymbol{\delta}}{ |\mathbf{x}|^3} + \frac{3 (\boldsymbol{\delta}\cdot \mathbf{x}) \mathbf{x}}{|\mathbf{x}|^5},\end{eqnarray*}
    where $\boldsymbol{\delta}$ is the doublet strength.

In the general body frame, the major axis of the prolate spheroid is along the $x$-axis. The equation of the spheroid rod is 
\begin{equation}
\frac{x^2}{a^2}+\frac{y^2+z^2}{b^2}=1,
\end{equation}
where $a$ is the major semi-major axis and $b$ is the semi-minor axes of the spheroid.  The related focal length $2c$ and the eccentricity $e$ are
\begin{equation}
c=\sqrt{{a^2}-{b^2}}= e a.
\end{equation}

From the results of Chwang \& Wu \cite{Wu1975}, by employing a line distribution of Stokeslets and potential doublets between the foci $x=-c$ and $c$, the velocity field of uniform flow $U_1 \mathbf{e}_x+U_2  \mathbf{e}_y$ past a prolate spheroid is 
\begin{eqnarray}\label{unifV}
&&\mathbf{u}(\mathbf{x})=U_1 \mathbf{e}_x+U_2  \mathbf{e}_y
-\int_{-c}^c \left[\alpha _1 \mathbf{u}_{\rm S} \left(\mathbf{x}-\bdxi ; \mathbf{e}_x\right)
+\alpha _2 \mathbf{u}_{\rm S} \left(\mathbf{x}-\bdxi , \mathbf{e}_y\right)\right] \, d\xi \\
&&+\int _{-c}^c\left(c^2-\xi ^2\right)\left[\beta _1\mathbf{u}_{\rm D}\left(\mathbf{x}-\bdxi ; \mathbf{e}_x\right)
+\beta _2\mathbf{u}_{\rm D}\left(\mathbf{x}-\bdxi , \mathbf{e}_y\right)\right]d\xi, \nonumber
\end{eqnarray}
in which the coefficients are
\begin{eqnarray*} 
\alpha _1=\frac{2 \beta _1e^2}{1-e^2}=\frac{U_1e^2}{-2 e+\left(1+e^2\right)L_e},\\
\alpha _2=\frac{2 \beta _2e^2}{1-e^2}=\frac{2 U_2e^2}{2 e+\left(3e^2-1\right)L_e},
\end{eqnarray*}
and 
\begin{eqnarray*} 
L_e&=&\log \left(\frac{1+e}{1-e}\right).
\end{eqnarray*}
The location of the singularities is $\boldsymbol{\xi}=\xi\mathbf{e}_x$.

For shear flow $\omega \,y\, \mathbf{e}_x$ past the spheroid, the velocity field is \cite{Wu1975}  
\begin{eqnarray}\label{shearVx}
\mathbf{u}(\mathbf{x})&=&\omega y \mathbf{e}_x+\int _{-c}^c\left(c^2-\xi ^2\right)\left[\alpha _3\mathbf{u}_{\rm SS}(\mathbf{x}-\bdxi ; \mathbf{e}_x,  \mathbf{e}_y)
\right.\nonumber\\
&&\left.+\gamma _3\mathbf{u}_{\rm R}\left(\mathbf{x}-\bdxi ; \mathbf{e}_z\right)\right]d\xi
+\beta _3\int _{-c}^c\left(c^2-\xi ^2\right)^2\frac{\partial }{\partial  y} \mathbf{u}_{\rm D}\left(\mathbf{x}-\bdxi ; \mathbf{e}_x\right)d\xi ,
\end{eqnarray}
where
\begin{eqnarray*}
\gamma _3&=&\omega \frac{1-e^2}{-2 e+\left(1+e^2\right)L_e},\\
\alpha _3&=&\frac{4e^2}{1-e^2}\beta _3=2 e^2\gamma _3\frac{-2e+L_e}{2 e\left(2e^2-3\right)+3\left(1-e^2\right)L_e}.
\end{eqnarray*}
For shear flow 
$\omega ' \,x \, \mathbf{e}_y$ past a spheroid, the velocity field is  
\begin{eqnarray}\label{shearVy}
\mathbf{u}(\mathbf{x})&=&\omega 'x  \mathbf{e}_y-\int _{-c}^c\left(c^2-\xi ^2\right)\left[\alpha _3'\mathbf{u}_{\rm SS}(\mathbf{x}-\bdxi ; \mathbf{e}_x, \mathbf{e}_y)
+\gamma _3'\mathbf{u}_{\rm R}\left(\mathbf{x}-\bdxi ; \mathbf{e}_z\right)\right]d\xi
\\&&
-\beta _3'\int _{-c}^c\left(c^2-\xi ^2\right)^2\frac{\partial }{\partial  x} \mathbf{u}_{\rm D}\left(\mathbf{x}-\bdxi ; \mathbf{e}_y\right)d\xi,\nonumber
\end{eqnarray}
where
\begin{eqnarray*}
\gamma _3'&=&\omega '\frac{1}{-2 e+\left(1+e^2\right)L_e},\\
\alpha _3'&=&\frac{4e^2}{1-e^2}\beta _3'= e^2\gamma _3'\frac{-2e+\left(1-e^2\right)L_e}{2 e\left(2e^2-3\right)+3\left(1-e^2\right)L_e}.
\end{eqnarray*}

Therefore, for a background flow \eqref{backgroudV} past a spheroid rod, the velocity field is the sum of velocities \eqref{unifV}, \eqref{shearVx} and \eqref{shearVy}
and we get \eqref{exact_vel} with
\begin{eqnarray*}
\chi _1=\alpha _3-\alpha _3',\quad 
\chi _2=\beta _3-\beta _3',\quad 
\chi _3=\left.(\gamma _3-\gamma _3'\right).
\end{eqnarray*}


\begin{thebibliography}{10}

\bibitem{balk2014kilohertz}
Andrew~L Balk, Lamar~O Mair, Pramod~P Mathai, Paul~N Patrone, Wei Wang, Suzanne
  Ahmed, Thomas~E Mallouk, J~Alexander Liddle, and Samuel~M Stavis.
\newblock Kilohertz rotation of nanorods propelled by ultrasound, traced by
  microvortex advection of nanoparticles.
\newblock {\em ACS nano}, 8(8):8300--8309, 2014.

\bibitem{Batchelor1970}
G.~K. Batchelor.
\newblock {Slender-body theory for particles of arbitrary cross-section in
  Stokes flow}.
\newblock {\em J. Fluid Mech.}, 44:419--440, 1970.

\bibitem{Blake1971}
J.~R. Blake.
\newblock A note on the image system for a stokelet in a no-slip boundary.
\newblock {\em Math. Proc. Camb. Philos. Soc.}, 70:303--310, 1971.

\bibitem{Buchmann2015}
Amy~L. Buchmann, Lisa~J. Fauci, Karin Leiderman, Eva~M. Strawbridge, and
  Longhua Zhao.
\newblock {\em Flow Induced by Bacterial Carpets and Transport of Microscale
  Loads}, pages 35--53.
\newblock Springer New York, New York, NY, 2015.

\bibitem{Camassa2011JFM}
R.~Camassa, R.~M. {McLaughlin}, and L.~Zhao.
\newblock Lagrangian blocking in highly viscous shear flows past a sphere.
\newblock {\em J. Fluid Mech.}, 669:120--166, feb 2011.

\bibitem{Wu1975}
A.~T. Chwang and T.~Y. Wu.
\newblock {Hydromechanics of low-Reynolds-number flow. Part 2. Singularity
  method for Stokes flows}.
\newblock {\em J. Fluid Mech.}, 67:787--815, 1975.

\bibitem{Cohen2006Suppressing}
A.~E. Cohen and W.~E. Moerner.
\newblock Suppressing brownian motion of individual biomolecules in solution.
\newblock {\em Proc. Natl. Acad. Sci. U.S.A.}, 103(12):4362--5, 2006.

\bibitem{Cortez2001}
R.~Cortez.
\newblock The method of regularized stokeslets.
\newblock {\em SIAM J.Sci. Comput.}, 23:1204, 2001.

\bibitem{cortez2005method}
Ricardo Cortez, Lisa Fauci, and Alexei Medovikov.
\newblock The method of regularized stokeslets in three dimensions: analysis,
  validation, and application to helical swimming.
\newblock {\em Phys. Fluids}, 17(3):031504, 2005.

\bibitem{dao2003mechanics}
Ming Dao, Chwee~Teck Lim, and Subra Suresh.
\newblock Mechanics of the human red blood cell deformed by optical tweezers.
\newblock {\em J. Mech. Phys. Solids}, 51(11):2259--2280, 2003.

\bibitem{davies1993mechanical}
Peter~F Davies and Satish~C Tripathi.
\newblock Mechanical stress mechanisms and the cell. an endothelial paradigm.
\newblock {\em Circ. Res.}, 72(2):239--245, 1993.

\bibitem{Du2003}
Q.~Du, M.~Gunzburger, and L.~Ju.
\newblock Constrained centroidal voronoi tessellations for surfaces.
\newblock {\em SIAM J. Sci. Comput.}, 24:1488--1506, 2003.

\bibitem{flierl1981particle}
Glenn~R Flierl.
\newblock Particle motions in large-amplitude wave fields.
\newblock {\em Geophys. Astrophys. Fluid Dyn.}, 18(1-2):39--74, 1981.

\bibitem{Haller2015Microfluidic}
Anna Haller, Andreas Spittler, Lukas Brandhoff, Helene Zirath, Dietmar
  Puchberger-Enengl, Franz Keplinger, and Michael~J. Vellekoop.
\newblock Microfluidic vortex enhancement for on-chip sample preparation.
\newblock {\em Micromachines}, 6(2):239--251, 2015.

\bibitem{hjelmfelt1966motion}
AT~Hjelmfelt and LF~Mockros.
\newblock Motion of discrete particles in a turbulent fluid.
\newblock {\em Appl. Sci. Res.}, 16(1):149--161, 1966.

\bibitem{kaftori1995particle}
D~Kaftori, G~Hetsroni, and S~Banerjee.
\newblock Particle behavior in the turbulent boundary layer. i. motion,
  deposition, and entrainment.
\newblock {\em Phys. Fluids}, 7(5):1095--1106, 1995.

\bibitem{Karimi2013}
A.~Karimi, S.~Yazdi, and A.~M. Ardekani.
\newblock Hydrodynamic mechanisms of cell and particle trapping in
  microfluidics.
\newblock {\em Biomicrofluidics}, 7(2):021501, 2013.

\bibitem{Lee2004Manipulation}
H~Lee, A.~M Purdon, and R.~M Westervelt.
\newblock Manipulation of biological cells using a microelectromagnet matrix.
\newblock {\em Appl. Phys. Lett.}, 85(6):1063--1065, 2004.

\bibitem{lee2010rapid}
Myung~Gwon Lee, Sungyoung Choi, and Je-Kyun Park.
\newblock Rapid multivortex mixing in an alternately formed
  contraction-expansion array microchannel.
\newblock {\em Biomed. Microdevices}, 12(6):1019--1026, 2010.

\bibitem{li2013chip}
Sixing Li, Xiaoyun Ding, Feng Guo, Yuchao Chen, Michael~Ian Lapsley,
  Sz-Chin~Steven Lin, Lin Wang, J~Philip McCoy, Craig~E Cameron, and Tony~Jun
  Huang.
\newblock An on-chip, multichannel droplet sorter using standing surface
  acoustic waves.
\newblock {\em Anal. Chem.}, 85(11):5468--5474, 2013.

\bibitem{Liu2016}
Zhihai Liu, Yunhao Chen, Li~Zhao, Yu~Zhang, Yong Wei, Hanyang Li, Yongjun Liu,
  Yaxun Zhang, Enming Zhao, Xinghua Yang, Jianzhong Zhang, and Libo Yuan.
\newblock Single-fiber tweezers applied for dye lasing in a fluid droplet.
\newblock {\em Opt. Lett.}, 41(13):2966--2969, Jul 2016.

\bibitem{mach2011automated}
Albert~J Mach, Jae~Hyun Kim, Armin Arshi, Soojung~Claire Hur, and Dino
  Di~Carlo.
\newblock Automated cellular sample preparation using a centrifuge-on-a-chip.
\newblock {\em Lab Chip}, 11(17):2827--2834, 2011.

\bibitem{Martindale2016}
J.~D. Martindale, M.~Jabbarzadeh, and H.~C. Fu.
\newblock Choice of computational method for swimming and pumping with
  nonslender helical filaments at low reynolds number.
\newblock {\em Phys. Fluids}, 28(2):021901, 2016.

\bibitem{Petit2012}
Tristan Petit, Li~Zhang, Kathrin~E. Peyer, Bradley~E. Kratochvil, and
  Bradley~J. Nelson.
\newblock Selective trapping and manipulation of microscale objects using
  mobile microvortices.
\newblock {\em Nano Lett.}, 12(1):156--160, 2012.

\bibitem{Pozrikidis1997}
C.~Pozrikidis.
\newblock {\em Introduction to theoretical and computational fluid dynamics}.
\newblock New York : Oxford University Press, 1997.

\bibitem{Renka1997}
R.~Renka.
\newblock Algorithm 772, stripack: Delaunay triangulation and voronoi diagrams
  on the surface of a sphere.
\newblock {\em ACM Trans. Math. Soft.}, 23:416--434, 1997.

\bibitem{stone2001}
H.~A. Stone and S.~Kim.
\newblock Microfluidics: basic issues, applications, and challenges.
\newblock {\em AIChE J.}, 47(6):1250--1254, 2001.

\bibitem{tanyeri2013manipulation}
Melikhan Tanyeri and Charles~M Schroeder.
\newblock Manipulation and confinement of single particles using fluid flow.
\newblock {\em Nano Lett.}, 13(6):2357--2364, 2013.

\bibitem{wang2013vortex}
Xiao Wang, Jian Zhou, and Ian Papautsky.
\newblock Vortex-aided inertial microfluidic device for continuous particle
  separation with high size-selectivity, efficiency, and purity.
\newblock {\em Biomicrofluidics}, 7(4):044119, 2013.

\bibitem{ye2012micro}
Zhou Ye, Eric Diller, and Metin Sitti.
\newblock Micro-manipulation using rotational fluid flows induced by remote
  magnetic micro-manipulators.
\newblock {\em J. Appl. Phys}, 112(6):064912, 2012.

\bibitem{ye2014dynamic}
Zhou Ye and Metin Sitti.
\newblock Dynamic trapping and two-dimensional transport of swimming
  microorganisms using a rotating magnetic microrobot.
\newblock {\em Lab Chip}, 14(13):2177--2182, 2014.

\bibitem{Zhao2010}
L.~Zhao.
\newblock {\em Fluid-Structure Interaction in Viscous Dominated Flows}.
\newblock PhD thesis, the University of North Carolina at Chapel Hill, July
  2010.

\bibitem{Zhao2015}
L.~Zhao.
\newblock Transportation particles in flows driven by nodal cilia processing
  {D}-shaped cones.
\newblock {\em J. Coupled Syst. Multiscale Dyn.}, 3(3):244--252, 2015.

\bibitem{Zhou2017Dumbbell}
Qi~Zhou, Tristan Petit, Hongsoo Choi, Bradley~J Nelson, and Li~Zhang.
\newblock Dumbbell fluidic tweezers for dynamical trapping and selective
  transport of microobjects.
\newblock {\em Adv. Funct. Mater.}, 27(1):1604571, 2017.

\end{thebibliography}

\bibliographystyle{plain}   

\end{document}